\documentclass[aps,epsfig,twocolumn,showpacs,superscriptaddress]{revtex4}
\usepackage{graphicx} 
\usepackage{epsfig} 
\usepackage{dcolumn}
\usepackage{color}
\newcommand{\bm}[1]{ \mbox{\boldmath $#1$}  }

\begin{document}

\title{Structure and three-body decay of $^9$Be resonances }

\author{R. \'Alvarez-Rodr\'{\i}guez} 
\email[electronic address: ]{raquel.alvarez@fis.ucm.es}
\affiliation{
  Department of Physics and Astronomy, University of Aarhus, \\
  DK-8000 Aarhus C, Denmark}
\affiliation{Grupo de F\'{\i}sica Nuclear. \\
  Departamento de F\'{\i}sica At\'omica, Molecular y Nuclear. \\
  Universidad Complutense de Madrid, \\ E-28040 Madrid, Spain}
\author{A.S. Jensen} \affiliation{
  Department of Physics and Astronomy, University of Aarhus, \\
  DK-8000 Aarhus C, Denmark} \author{E. Garrido}
\affiliation{Instituto de Estructura de la Materia,
  Consejo Superior de Investigaciones Cient\'{\i }ficas, \\
  Serrano 123, E-28006 Madrid, Spain} \author{D.V. Fedorov}
\affiliation{
  Department of Physics and Astronomy, University of Aarhus, \\
  DK-8000 Aarhus C, Denmark}
\date{\today}

\begin{abstract}
The complex-rotated hyperspherical adiabatic method is used to study
the decay of low-lying $^9$Be resonances into one neutron and two
$\alpha$-particles.  We investigate the six resonances above the
break-up threshold and below 6 MeV: $1/2^\pm$, $3/2^\pm$ and $5/2^\pm$.
The short-distance properties of each resonance are studied, and the
different angular momentum and parity configurations of the $^8$Be and
$^5$He two-body substructures are determined.  We compute the
branching ratio for sequential decay via the $^8$Be ground state which
qualitatively is consistent with measurements.  We extract the
momentum distributions after decay directly into the three-body
continuum from the large-distance asymptotic structures.  The
kinematically complete results are presented as Dalitz plots as well
as projections on given neutron and $\alpha$-energy. The distributions
are discussed and in most cases found to agree with available
experimental data.
\end{abstract}


\maketitle

\section{Introduction} 

The structure of $^9$Be has been extensively studied, both
theoretically
\cite{kun60,tan62,hiu63,bar66,bou68,gru70,zah76,fon88,fur80,kan95,ara96,pie02,
  gri02,ara03,for05,ibr09} and experimentally
\cite{moe64,chr66,che70,jer78,nym90,boc90,sum02,ful04,pre04,pap07,bro07}, but 
there are still large uncertainties in the structure and decay of the
low-lying excited states.  This is surprising and worrisome in view of
the large efforts and the expected rather accurate approximation as a
simple three-body system where the intrinsic degrees of freedom are
inactive.  Are the problems related to inaccuracies of the theoretical
models, the numerical techniques, direct experimental uncertainties,
data analysis, or interpretation of the data in comparison with model
results?

In theory the three-body continuum problem is better handled and more
accurately solved for nuclear systems with the special mixture of
short and long-range interactions.  Observables rather close to the
directly measured quantities can be delivered. In experiments both
beam quality, detector systems and systematic analyses have improved
substantially in recent years. This means that genuine three-body
systems can be treated fully and precisely in both theory and
experiment, more specifically complete kinematics of the fragments are
available.  The road to detailed comparison is therefore paved. The
simplest systems should then be understood before reliability can be
expected for more complicated scenarios.

Furthermore, the results of requested applications in astrophysics, 
where often the energies are too low to be reached experimentally, can
only be indirectly tested by their implications. The
approximations employed so far in predictions should then be tested by
comparison.  A reasonable procedure is to select a three-body system,
compute and measure the best we can, and compare as detailed as
possible.  The choice of $^9$Be is tempting as a rather simple system
which is accessible to both theory and experiments. In addition, this
is a system of particular interest in astrophysics, where formation of
$^9$Be can proceed through the reaction $\alpha$($\alpha
n$,$\gamma$)$^9$Be.  The subsequent reactions,
$^9$Be($\alpha$,$n$)$^{12}$C, link to heavier elements in stellar
nuclear synthesis responsible for the present Universe.

From the early days of Nuclear Physics, the structure of the
$^9$Be nucleus has been considered a prototype of the cluster-like
structure of nuclei. Therefore, different types of three-body
descriptions have been used to describe it: early cluster models
\cite{kun60,tan62,hiu63,gru70,fur80}, and more sophisticated ones,
e.g. the Resonating Group Model \cite{zah76}, Antisymmetrized
Molecular Dynamics \cite{kan95}, or the Microscopic Multicluster Model
\cite{ara03}. Moreover, many-body type of calculations have also been
performed on $^9$Be: projected Hartree-Fock \cite{bou68}, Shell Model
\cite{bar66}, Quantum Monte Carlo \cite{pie02}, and {\em ab initio}
no-core shell model \cite{for05}. All of them are able to reproduce
the low-lying energy spectrum and electromagnetic properties in fair
agreement with the experimental data available at the moment, though
in general theoretical models predict more states than are seen
experimentally.

\begin{figure}
\begin{center}
\vspace*{-0.2cm}
\epsfig{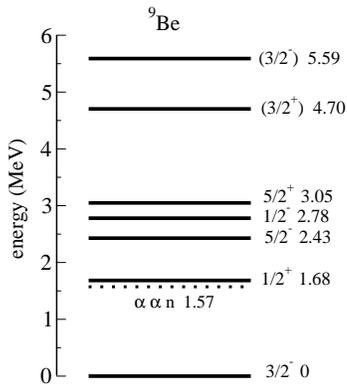}
\end{center}
\vspace*{-0.1cm}
\caption{Scheme of the experimentally known levels of $^9$Be below 6~MeV
of excitation energy. }
\label{figlev}
\end{figure}

Somewhat surprisingly, the three-body decays of the $^9$Be resonances
have been barely studied \cite{gri02,alv08}. The inverse process may
proceed through the resonances but non-resonant contributions are also
important. Before facing this more complicated process, it is
advisable to get a good understanding of the resonance decay of $^9$Be
into $\alpha \alpha n$.  The experimentally known $^9$Be states are
shown in Fig.~\ref{figlev} for excitation energies below 6~MeV where
all other particle thresholds than $\alpha \alpha n$ are closed.  All
these levels, apart from $\frac{5}{2}^-$, have a fairly large width,
which makes it difficult to determine their properties.  Many
experimental efforts have been addressed towards this
$\frac{5}{2}^-$-state \cite{moe64,che70,boc90,pap07}.  They all agree
in the small percentage of the decay taking place via the $^8$Be
ground state.  So far no agreement has been reached regarding its main
decay path, via $^8$Be($2^+$), $^5$He($p$), or direct.  Much less is
known about the decays of other low-lying resonances of $^9$Be. Both,
the $\frac{1}{2}^-$ and $\frac{5}{2}^+$, seem to prefer to decay
through $^8$Be($0^+$), although especially the results for
$\frac{1}{2}^-$ need to be better established.

The purpose of the present article is to report on comprehensive
calculations of the three-body properties of low-lying states in
$^9$Be.  We give a survey of the short-distance structure of the
resonances, their dynamic evolution across intermediate distances
which often is referred to as decay mechanism, and eventually reaching
the large-distance asymptotics which reveal the complete set of
momentum distributions of the fragments after decay.  The
two-dimensional energy correlations showed in Dalitz plots can be
directly compared to the experimental data. This is the only
information relating measurements with initial short-distance
structure and decay mechanism. Extrapolations backward from data,
therefore necessarily must be model dependent. We attempt to provide
an interpretation which is as physically meaningful as possible.

\section{Theoretical ingredients}

\begin{figure}
\begin{center}
\vspace*{0.1cm}
\epsfig{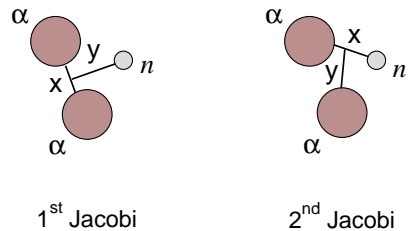}
\end{center}
\vspace*{-0.1cm}
\caption{(Color online) Scheme of the different Jacobi coordinates for
  $^9$Be. }
\label{figjac}
\end{figure}

The decay of $^9$Be into two $\alpha$-particles and one neutron is
obviously a three-body problem in the final state where the particles
are far from each other.  Furthermore the dominant structure at small
distances is also of cluster nature for these low-lying resonances.
The Hamiltonian for this cluster structure is then 
\begin{eqnarray}
 H = \sum_{i=}^{3} \frac{\bm{p}^2_i}{2m_i} - \frac{\bm{P}^2_t}{2M} +
 \sum_{i=1}^{3} V_i(\bm{r}_j - \bm{r}_k) + V_{3b} \; ,
\label{e27}
\end{eqnarray}
where $\bm{p}_i$ and $m_i$ are momentum and mass of particle $i$,
$\bm{P}_t$ is the total momentum, $V_i$ is interaction between
particle $j$ and $k$. Here \{$i,j,k$\} is a cyclic permutation of
\{$1,2,3$\},  and $V_{3b}$ is a three-body potential depending on all 
three particle coordinates.  It is convenient to substitute the
position coordinates by the Jacobi coordinates, $\bm{x}$ and
$\bm{y}$ defined as:
\begin{eqnarray}
\bm{x}&=&\sqrt{\frac{\mu_{12}}{m}}
(\bm{r}_{1}-\bm{r}_{2}) \nonumber \\
\bm{y}&=&\sqrt{\frac{\mu_{12,3}}{m}}
\left( \bm{r}_{3}-\frac{m_{1}\bm{r}_{1}+m_{2}\bm{r}_{2}}
{m_{1}+m_{2}} \right) ,
\label{eq4}
\end{eqnarray}
where $m$ is an arbitrary mass scale chosen as the nucleon mass, and
$\mu_{12}$ and $\mu_{12,3}$ are the reduced masses.  The Hamiltonian
becomes
\begin{eqnarray}
 H = - \frac{\hbar^2}{2m}(\bm{\nabla}^2_x + \bm{\nabla}^2_y)  +
 \sum_{i=1}^{3} V_i(\bm{x},\bm{y}) + V_{3b} \; . \label{e29}
\end{eqnarray}
In the present case we have two possible choices for Jacobi
coordinates (see Fig.~\ref{figjac}), leading to different sets of
($\bm{x},\bm{y}$)-coordinates.  We use hyperspherical coordinates
where the six coordinates are
\{$\rho$,$\alpha$,$\theta_x$,$\phi_x$,$\theta_y$,$\phi_y$\}. The
$\theta$'s and $\phi$'s refer to the directions of $\bm{x}$ and
$\bm{y}$, while $\alpha = \arctan (x/y)$ and $\rho = (x^2+y^2)^{1/2}$
are related to their sizes. Actually $\rho$ is the only length
coordinate, which describes the average distance from the center of
mass.

Resonances are computed within the formalism of complex scaling of
the hyperspherical coordinates. This is particularly simple since only
one coordinate, the hyperradius $\rho$, has to be scaled. The
Hamiltonian is complex-rotated, i.e.
\begin{equation}
H_\theta (\rho) = H (\rho e^{i\theta}).
\end{equation}
We use the adiabatic expansion method and solve the Faddeev equations
stepwise, that is first the angular then the (hyper) radial part
\cite{nie01}.  The angular part of the Hamiltonian is first solved
keeping fixed the value of $\rho$, i.e.
\begin{equation}
T_\Omega \Phi_{nJM}^{(i)}+\frac{2m}{\hbar^2} \rho^2V_i
\Phi_{nJM}  = \lambda_n \Phi_{nJM}^{(i)} \qquad i=1,2,3\;,
\label{fadang}
\end{equation}
where $n$ labels the adiabatic components. $T_\Omega$ is the angular
part of the kinetic energy operator \cite{nie01}. This provides a
complete set of angular wave-functions, $\Phi_{nJM}$, that are
employed to expand the total wave-function $\Psi^{JM}$:
\begin{equation}
\Psi^{JM} = \frac{1}{\rho^{5/2}}\sum_n f_n (\rho) \Phi_{nJM}
(\rho,\Omega)\;,
\end{equation}
where the $\rho$-dependent expansion coefficients, $f_n (\rho)$, are
the hyperradial wave functions obtained from the coupled set of
hyperradial equations \cite{nie01}:
\begin{eqnarray}
  \lefteqn{\hspace*{-1.cm}
    \left[ -\frac{d^2}{d\rho^2} +  \frac{2m}{\hbar^2} 
      (V_{\mbox{\tiny eff}}^{(n)}(\rho)+V_{3b}(\rho) - E)
    \right] f_n(\rho) }
  \nonumber \\&&
  - \sum_{n'} \left( 2 P_{n n'} \frac{d}{d\rho} + Q_{n n'} \right)f_{n'}(\rho)
  = 0 
\label{radial}
\end{eqnarray}
where $V_{3b}$ is a three-body potential used for fine-tuning and
the functions $P_{n n'}$ and $Q_{n n'}$ are given for instance in
\cite{nie01}. The eigenvalues $\lambda_n$ in Eq.(\ref{fadang})
enter in (\ref{radial}) as a part of the effective adiabatic potentials:
\begin{equation}
V_{\mbox{\tiny eff}}^{(n)}(\rho)=\frac{\hbar^2}{2 m}\frac{1}{\rho^2}
                               \left( \lambda_n(\rho)+\frac{15}{4} \right).
\label{adpot}
\end{equation}

Resonances are usually understood as states with complex energy
$E=E_R- i \Gamma_R/2$, where $E_R$ is the energy of the resonance and
$\Gamma_R$ is the width. If we define now $k=\sqrt{2\mu |E| /
  \hbar^2}$, with $\mu$ being the reduced mass of the system, we then
have that the asymptotic form of the resonance wave function is given
by
\begin{equation}
  f_n(\rho \to \infty) \sim e^{i\:  k\rho \cos \beta} e^{k\rho\sin \beta}\;,
\end{equation}
where with $\beta=\frac{1}{2}\arctan(\frac{\Gamma_R}{2E_R})$. The
first term oscillates while the second one diverges. After the complex
scaling transformation ($\rho \to \rho e^{i\theta}$) the radial
asymptotic behavior becomes
\begin{equation}
f_n(\rho \to \infty)
\sim e^{i\: k\rho \cos (\theta-\beta)} e^{-k\rho\sin (\theta-\beta)} \;,
\end{equation}
which implies that when $\theta > \beta$ the wave function goes to
zero exponentially and the resonance can then be obtained as an
ordinary bound state. True bound states remain unchanged under the
coordinate rotation.

For our particular case of $^9$Be, the two-body interactions, $V_i$,
are chosen to reproduce the low-energy scattering properties of the
two different pairs of particles in our three-body system. We use the
Ali-Bodmer $\alpha-\alpha$ potential \cite{ali66} supplemented by the
Coulomb potential between $\alpha$-particles, and the $\alpha$-neutron
interaction is taken from \cite{cob97}.  The $^{9}$Be-resonances are
of three-body character at large-distances, since no other channels
are open for these energies.  This is not necessarily correct at
short-distances where all 9 nucleons (and their intrinsic structure)
may contribute in different (cluster) configurations.

We use the (complex scaled) three-body model at all distances because
the decay properties only require the proper description of the
emerging three particles. Therefore, the angular eigenfunctions and
eigenvalues in Eq.(\ref{fadang}) are complex, as well as all the terms
entering in the coupled set of radial equations (\ref{radial}). The
missing information, if any, beyond the three-body structure, is the
initial structure at small distances. This piece, acting as a boundary
condition, is parametrized through a short-range three-body potential
of the form $V_{3b} = S\exp(-\rho^2/b^2)$. 

Different three-body resonances correspond in general also to
different three-body structures. As a consequence, the missing
information going beyond the two-body correlations is in principle
resonance dependent. The strength (and possibly also the range) in the
three-body force is therefore adjusted individually to give the
correct position of each of the resonance energies. This adjustment
implies that the potential is angular momentum dependent but this is
already a property of the two-body potential. The corresponding
Hamiltonian for the three-body problem still exists as a non-local
operator, but this feature is already present due to the angular
momentum dependence of the two-body interactions. It is then clear
that this phenomenological fine-tuning is not arising from the
presence of a genuine three-body interaction.  

The energy dependence is all-decisive for decay properties as evident
in the exponential dependence of probability for tunneling through a
barrier. On the other hand, the three-body potential is assumed to be
completely structure independent, and therefore only marginally
influencing the partition between different structures at large
distances. However, this is an assumption which may be violated
through the dynamic evolution from inaccurate initial small-distance
boundary conditions provided by the three-body potential.

\section{Short-distance structure}

The short-distance structures are crucial for the energies whereas
dominating configurations at large distances are decisive for the
observable decay properties. The connection between these two regimes
contain information about the decay mechanism which therefore only is
an observable effect precisely to the extent reflected in the final
distributions.  In other words, sensible theoretical models are
indispensable to interpret the experimental results.  In this section
we extract and discuss short-distance bulk properties, that is
effective potentials, energies and partial-wave structure.

\subsection{Adiabatic potentials and energies}
\label{sec3a}

\begin{figure}
\begin{center}
\epsfig{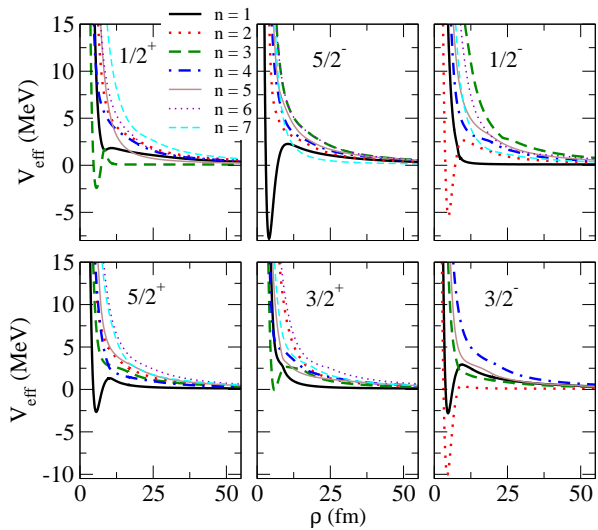}
\end{center}
\vspace*{-0.2cm}
\caption{(Color online) Real parts of the seven lowest adiabatic
  potentials as functions of the hyperradius $\rho$ for the $1/2^+$,
  $5/2^-$, $1/2^-$, $5/2^+$, $3/2^+$ and $3/2^-$ low-lying resonances
  in $^9$Be. }
\label{figpot}
\end{figure}

Each of the adiabatic potentials entering in Eq.(\ref{radial})
corresponds to a specific combination of quantum numbers,
i.e. partial-wave angular momenta between the particles in the
different Jacobi systems.  Usually only rather few adiabatic
potentials are needed to achieve convergence.  We show in
Fig.~\ref{figpot} the real part of these adiabatic potentials as
defined in (\ref{adpot}) plus the three-body potential individually
fitted for each spin and parity in order to reproduce the experimental
resonance energies.

We did not include the non-adiabatic diagonal parts in the figure
($Q_{nn}$ in (\ref{radial}), $P_{nn}=0$) because they usually are
insignificant and in a sense more related to the coupling between the
different radial potentials. The only exception is the deepest
potential for the 1/2$^+$ resonance. In this case the $Q_{33}$ term is
included, since this term is responsible for the potential barrier
that permits to hold the resonance (see \cite{gar10} for details). The
imaginary parts of the potentials are small, oscillate, and go to zero
for large values of $\rho$. They are mostly related to the widths.

In Fig.~\ref{figpot} we observe that at small distances the lowest
potentials have a pronounced well followed by a potential barrier that
are responsible for the bound states and the resonances. We have two
attractive potentials for angular momentum and parity $J^\pi = 3/2^-$;
the deepest one supports the ground state (bound) while the other one
supports the higher resonance.  For other $J^\pi$ values only the
lowest potential exhibits an attractive region at small distance. All
the other potentials are repulsive at all distances.

At large distances, the lowest potential is in all the cases 
approaching the $^8$Be($0^+$) resonance energy of $\sim$~0.1~MeV. Its
angular structure corresponds to the two alpha particles populating the
 $^8$Be($0^+$) resonance, while the
remaining neutron is far away and described through the radial
equation.  These specific potentials are labeled as number $n=3$ in
$1/2^+$, $n=7$ in $5/2^-$, $n=1$ in $1/2^-$, $5/2^+$ and $3/2^+$, and
$n=2$ in $3/2^-$.  They characterize then a decay mechanism where the
neutron first is emitted and this two-body $\alpha-\alpha$ resonance
is populated and subsequently decaying.

\begin{table}
  \caption{Calculated and measured energies $E_R$ (in MeV) and widths 
    $\Gamma_R$ (in MeV) of the $^9$Be resonances for different $J^\pi$. 
    The experimental 
    values (labeled ``exp'') are from \cite{web} and the calculated results 
    (labeled ``th'') are obtained with the three-body interaction parameter 
    $S$ (in MeV) (the range is taken $b=5$ fm in all the cases). 
    The energies are measured from the 
    $\alpha\alpha n$ threshold. }
\label{tab}
\vspace*{0.2cm}
\begin{tabular}{|cccccc|}
\hline
$J^\pi$&$E_{R,exp}$&$\Gamma_{R,exp}$&$E_{R,th}$&$\Gamma_{R,th}$&$S$\cr
\hline
$3/2^-$ & $-1.574$ & 0.0  & -1.60 & 0.0  & 2.5   \\
$1/2^+$ & $0.110\pm 0.020$ & $0.214\pm 0.005$ & 0.11 & $\simeq$ 0.1 & --  \\

$5/2^-$ & $0.855\pm0.013$ & $(7.8\pm 1.3)\cdot 10^{-4}$ & 0.86 & $7.0\times10^{-4}$ 
&3.7\\

$1/2^-$ & $1.21\pm0.12$ & $1.10\pm 0.12$ & 1.25 & 0.65 & 2.0 \\

$5/2^+$ & $1.475\pm 0.009$ & $0.282\pm 0.011$ & 1.46 & 0.34 & 0.7 \\

$3/2^+$ & $3.130\pm 0.025$ & $0.743\pm 0.055$ & 3.12 & 1.74 & 1.0 \\

$3/2^-$ & $4.02\pm 0.01$ & $1.33\pm 0.36$ & 2.65 & 0.93 & 2.5   \\
\hline
\hline
\end{tabular}
\end{table}

The complex scaling of the hyperradius leads to a Hamiltonian with
complex solutions vanishing exponentially at large distances precisely
as ordinary bound state wave functions.  The real and imaginary parts
of the complex three-body energy are, respectively, the resonance
energy $E_R$, and $-\Gamma_R/2$, where $\Gamma_R$ is the width of the
resonance. The computed results are collected in table~\ref{tab}
together with the known experimental values \cite{web}.  The
three-body strength is adjusted to give the correct energy position
for all $J^\pi$ except for $ 3/2^-$, where we choose the same values
for both bound state and resonance as in ref.~\cite{alv07a}.

The energies are given relative to the $\alpha-\alpha-n$ breakdown
threshold, at 1.57~MeV above the ground state. We have found one bound
state, which corresponds to the $^9$Be ground state, and six
resonances below 6~MeV of excitation energy.  The ground state has
$J^\pi = 3/2^-$ and the resonances $1/2^\pm$, $3/2^\pm$ and
$5/2^\pm$. Those states are most likely to contribute to processes
bridging the $A=5,8$ instability gaps in nuclear synthesis in suitable
astrophysical environments \cite{die10}.  We keep the range of the
three-body potential at $b=5.0$~fm for all $J^\pi$, while adjusting the
strength to place the resonance energies (and bound state) at the
desired measured position.  Thus, we did not attempt to reproduce the
widths.

Several features are interesting in table~\ref{tab}. First the
three-body potentials have very moderate strengths which were found to
reproduce the real part of the measured energies. This is fine-tuning
and indicates strongly that the dominating structures in fact really
are three-body clusters. It is then significant that all widths,
except for $1/2^{\pm}$ and the very narrow 5/2$^-$ state, are larger
than the corresponding measured values. This is consistent with a
many-body configuration at small distances which would decrease the
branching ratio of decay into the investigated three-body cluster
structure. A possible quantification of this deviation is in terms of
preformation factors expressing that only part of the complete
wave-function describes the three-body cluster. Again the deviation
amounts to about factors of two in agreement with only smaller
contributions from a many-body structure.  A smaller range compensated
by a slightly larger strength to leave the energy untouched would
decrease the width towards the measured values.

However, the 3/2$^-$ state is another exception in
table~\ref{tab}. The computed value of width is smaller than the
experimental table value. The computed energy is also below
measurements by about $1.4$~MeV. An attempt to increase the computed
energy by this amount with use of a repulsive three-body potential of
range $5$~fm immeduately cause the resonance to disaapear into the
continnum corresponding to an energy above the barrier. This shows
that the width in the model very quickly becomes very large, and
exceeding the experimental table value.  Either the model is missing
an important ingredient for this state or its width should be
substantially larger.

The $1/2^{\pm}$ states are an apparent exception in table~\ref{tab}
where the experimental widths are larger than the calculated values.
For $1/2^{+}$ this discrepancy has caused a good deal of trouble. In a
recent investigation \cite{gar10} this conundrum was explained as a
genuine three-body effect where the resonance structure changes from
dominantly $^5$He plus an $\alpha$-particle at small distances to
$^8$Be plus a neutron at large distances. The large measured width is
in fact obtained from an assumption of two-body character and
sequential decay in the $R$-matrix analysis of the photo dissociation
cross section \cite{sum02}. It is remarkable that a much smaller width
consistent with \cite{gar10} was obtained already many years ago
\cite{fur80}. The large experimental width of $1/2^{-}$ should be
reevaluated since we suspect that the $R$-matrix parametrization and
the sequential decay channel is used too strongly in the extraction
from the data analysis. This was argued for $1/2^{+}$ in \cite{gar10}.

\subsection{Partial waves}

The different two-body components of the three-body system are
constrained by the total angular momentum and parity of each
state. For example, in the first Jacobi (see Fig.~\ref{figjac}) the
two $\alpha$-particles must couple to an even orbital angular momentum
$\ell_x$. A neutron with even (odd) angular momentum, $\ell_y$, will
give a positive-parity (negative-parity) state. The orbital angular
momentum $\ell_y$ couples to the spin of the neutron to the angular
momentum $j_y$.  In the second Jacobi set $\ell_x$ can be either even
or odd, and couples to the spin of the neutron to the angular momentum
$j_x$.

We choose our partial-wave components taking into account these
selection rules. In the present case convergence is achieved with a
number of partial waves between 10 and 30, depending on the
resonance. The accuracy is optimized by choosing a large value for the
hypermomentum $K$ for the large contributions. Unfortunately, the
higher the value of $K_{max}$, the larger becomes the total number of
basis states. Therefore, $K_{max}$ must be chosen carefully for each
partial wave, trying to achieve accuracy while keeping the number of
basis elements as small as possible.

Tables \ref{par} and \ref{par2} show, for the first and second Jacobi
sets, the contributions $W$ to the total $J^\pi$ wave functions from
those components contributing more than 1\%. These contributions are
well defined for the complex rotated resonance wave function, since it
behaves asymptotically like a bound state.  The maximum value of the
hypermomentum $K$ is also given for each component.  The computed
values of $W$ are given in the last column of the tables. The wave
functions are located at relatively small distances, and the
contribution from the different components (obtained after integration
of the square of the wave function over all the hyperangular
variables) contain therefore information mainly about these bulk
structures.  The decay properties are contained in the large-distance
tails, whose partial wave content can be entirely different, as
discussed in details in the next section.

\begin{table}
  \caption{Components included for each $J^\pi$ state of $^9$Be relative
    to the first Jacobi set (see Fig.~\ref{figjac}). $q$ labels the set of 
    quantum numbers, $\ell_x$, $\ell_y$ and $j_y$ (coupling between $\ell_y$
and the spin of the neutron) are the angular momenta
    relative to $x$ and $y$ Jacobi coordinates. $K_{max}$ is the maximum 
    value of the hypermomentum and $W$ gives the
    probability in \% for finding these components in the resonance.
    Only the components contributing more than 1~\% are shown. }
\label{par}
\vspace*{0.2cm}
\begin{tabular}{|c|c|cccc|c|}
\hline
$J^\pi$         & $q$ & $\ell_x$ & $\ell_y$ & $j_y$    &$K_{max}$&$W$\\  
\hline
$\frac{1}{2}^+$ & 1 & 0     & 0    &$\frac{1}{2}$&150& 100\%  \\
\hline
$\frac{5}{2}^-$ & 1 & 0     & 3    &$\frac{5}{2}$&95& 1\%\\
                & 2 & 2     & 1    &$\frac{1}{2}$&50& 11\%\\
                & 3 & 2     & 1    &$\frac{3}{2}$&75& 83\%\\
                & 4 & 2     & 3    &$\frac{7}{2}$&50& 2\%\\
\hline
$\frac{1}{2}^-$ & 1 & 0     & 1    &$\frac{1}{2}$&220& 42\%\\
                & 2 & 2     & 1    &$\frac{3}{2}$&180& 55\%\\
                & 3 & 4     & 3    &$\frac{7}{2}$&150& 2\% \\
\hline
$\frac{5}{2}^+$ & 1 & 0     & 2    &$\frac{5}{2}$&125& 52\%\\
                & 2 & 2     & 0    &$\frac{1}{2}$&120& 27\%\\
                & 3 & 2     & 2    &$\frac{5}{2}$&100& 12\%\\
                & 4 & 2     & 4    &$\frac{9}{2}$&60& 4\%\\
\hline
$\frac{3}{2}^+$ & 1 & 0     & 2    &$\frac{3}{2}$&175& 10\%\\
                & 2 & 2     & 0    &$\frac{1}{2}$&155& 73\%\\
                & 3 & 2     & 2    &$\frac{3}{2}$&85& 4\%\\
                & 4 & 2     & 2    &$\frac{5}{2}$&35& 9\%\\
                & 5 & 4     & 2    &$\frac{5}{2}$&35& 3\%\\
\hline
$\frac{3}{2}^-$ & 1 & 0     & 1    &$\frac{3}{2}$&170& 3\%\\
                & 2 & 2     & 1    &$\frac{1}{2}$&70& 51\%\\
                & 3 & 2     & 1    &$\frac{3}{2}$&80& 41\%\\
                & 4 & 2     & 3    &$\frac{7}{2}$&40& 3\%\\
\hline
\end{tabular}
\end{table}

\begin{table}
  \caption{The same as table~\ref{par} for the second Jacobi set (see Fig.~\ref{figjac}).
The angular momentum $j_x$ results from the coupling between $\ell_x$ and the spin of the
neutron.}
\label{par2}
\vspace*{0.2cm}
\begin{tabular}{|c|c|cccc|c|}
\hline
$J^\pi$         & $q$ & $\ell_x$ & $j_x$ & $\ell_y$ &$K_{max}$&$W$  \\  
\hline
$\frac{1}{2}^+$ & 1 & 0 & $\frac{1}{2}$ &  0    &150& 50\%   \\  
                & 3 & 1 & $\frac{3}{2}$ &  1    &89&  50\%   \\  
\hline
$\frac{5}{2}^-$ & 1 & 1 &$\frac{1}{2}$ &  2    &50& 11\%\\
                & 2 & 1 &$\frac{3}{2}$ &  2    &70& 73\%\\
                & 3 & 2 &$\frac{5}{2}$ &  1    &30& 9\% \\
                & 4 & 2 &$\frac{5}{2}$ &  3    &65& 1\% \\
\hline
$\frac{1}{2}^-$ & 1 & 0 & $\frac{1}{2}$ &  1    &150& 4\% \\
                & 2 & 1 & $\frac{1}{2}$ &  0     &200& 35\%\\  
                & 3 & 1 & $\frac{3}{2}$ &  2    &200& 50\%\\  
                & 4 & 2 & $\frac{3}{2}$ &  1     &150& 6\% \\  
                & 5 & 2 & $\frac{5}{2}$ &  3     &120& 1\% \\  
\hline
$\frac{5}{2}^+$ & 1 & 0 & $\frac{1}{2}$ &  2     &95& 1\% \\
                & 2 & 1 & $\frac{1}{2}$ &  3     &95& 5\% \\
                & 3 & 1 & $\frac{3}{2}$ &  1     &125& 50\% \\
                & 4 & 1 & $\frac{3}{2}$ &  3     &95& 5\% \\
                & 5 & 2 & $\frac{3}{2}$ &  2     &95& 1\% \\
                & 6 & 2 & $\frac{5}{2}$ &  0     &95& 25\% \\
\hline
$\frac{3}{2}^+$ & 1 & 0 & $\frac{1}{2}$ &  2     &99& 27\% \\  
                & 2 & 1 & $\frac{1}{2}$ &  1     &99& 21\% \\ 
                & 3 & 1 & $\frac{3}{2}$ &  1     &55& 12\% \\  
                & 4 & 1 & $\frac{3}{2}$ &  3     &99& 21\% \\ 
                & 5 & 2 & $\frac{3}{2}$ &  0     &25& 6\% \\
                & 6 & 2 & $\frac{5}{2}$ &  2     &35& 3\% \\
\hline
$\frac{3}{2}^-$ & 1 & 0 & $\frac{1}{2}$ &  1     &60& 1\% \\   
                & 2 & 1 & $\frac{1}{2}$ &  2     &50& 45\% \\  
                & 3 & 1 & $\frac{3}{2}$ &  0     &95& 3\% \\  
                & 4 & 1 & $\frac{3}{2}$ &  2     &90& 34\% \\  
                & 5 & 2 & $\frac{3}{2}$ &  1     &50& 6\% \\ 
                & 6 & 2 & $\frac{5}{2}$ &  1     &50& 5\% \\ 
\hline
\end{tabular}
\end{table}

The lowest resonance, $J^{\pi}=\frac{1}{2}^+$, is located only
$18$~keV above the two-body $^8$Be narrow ground state resonance at
$918$~keV. In the first Jacobi coordinates this state is entirely
described as $s$-waves between the $\alpha$-particles and therefore
also between their center of mass and the neutron.  The interesting
structure is seen in the two other identical Jacobi coordinates where
the structure changes abruptly from $\alpha-$neutron $p_{3/2}$ to
$s_{1/2}$ configurations at around $10$~fm, see \cite{gar10}.  The
bulk part of the resonance structure found at small distances then
roughly amounts to equal parts in each of these partial waves.

The next resonance with $J^{\pi}=\frac{5}{2}^-$ is very narrow due 
to the large barrier in the dominating partial
wave of $(\ell_x,\ell_y)=(2,1)$ in the first Jacobi and $(1,2)$ in the
second set of Jacobi coordinates.  This can be described as
$^8$Be($2^+$) or $^5$He($p_{3/2}$), respectively, but it is in fact the
same state in different coordinate systems. It is therefore not
meaningful to distinguish between these configurations unless also
spatial distributions are included in the distinction \cite{alv08}.

The $\frac{1}{2}^-$ resonance is a result of the
$^5$He  $p$-wave  attraction combined with orbital angular momentum
coupling to $2$ of the last $\alpha$-particle. Only the
corresponding adiabatic potential is really attractive. This
configuration translates to $(\ell_x,\ell_y)=(0,1),(2,1)$ in the first
Jacobi coordinates where only even $\ell_x$ are allowed.

The $\frac{5}{2}^+$ resonance is dominated by a combination of
$^5$He($p_{3/2}$) and $^8$Be($0^+,2^+$). Only one of the adiabatic
potentials is really attractive and in fact not very deep. This state
is important at moderate temperatures for photo dissociation and
three-body recombination from the continuum via $E1$-transitions
\cite{die10}.

The next resonance, $\frac{3}{2}^{+}$, is higher.  Its structure is
similar to the $\frac{5}{2}^{+}$ state in the first Jacobi where
$(\ell_x,\ell_y)=(0,2),(2,0)$ are roughly interchanged in the two
states. In the second Jacobi system the $^5$He($p_{3/2}$) structure
also has a relevant, although not dominant, contribution.

The last resonance, $\frac{3}{2}^-$, has $(\ell_x,\ell_y)=(2,1)$ and
$(1,2)$ in the first and second Jacobi sets, respectively. This is
reflecting a combination of the influence of the interactions related
to the $^5$He($p_{3/2}$) and $^8$Be($2^+$) two-body resonances.  The
similarity to the $\frac{5}{2}^-$ state is striking, except for the
larger width arising from a higher excitation energy.

\section{Long-distance structure}

Resonances may be populated at small distances via beta-decay or some
specific reactions, but the products after the resonance decay reflect
the behavior at large distances. The short and large-distance
structures are related through the quantum mechanical solution, and
the configurations sometimes change dramatically with the
hyperradius. This connection from small to large distances is
therefore crucial for the interpretation of the decay mechanism and
the measured results. We shall first show the dynamic evolution of
each resonance configuration, and afterwards show the momentum
distributions of the fragments as Dalitz plots with the full
information.

\subsection{Dynamic evolution}

As mentioned in section \ref{sec3a}, at large distances, the lowest
adiabatic potential is, for all the resonances, approaching the
$^8$Be($0^+$) resonance energy of $\sim$~0.1~MeV. Its angular
structure corresponds to the two alpha particles populating the
$^8$Be($0^+$) resonance, while the remaining neutron is far away. In
other words, for large values of $\rho$, the configuration of this
potential in the first Jacobi set approaches $\ell_x=0$, $j_y=J$, and
$\ell_y$ has to be one of the $J \pm 1/2$ values in order to produce
the correct parity.

\begin{figure}[th!]
\begin{center}
\vspace*{0.2cm}
\epsfig{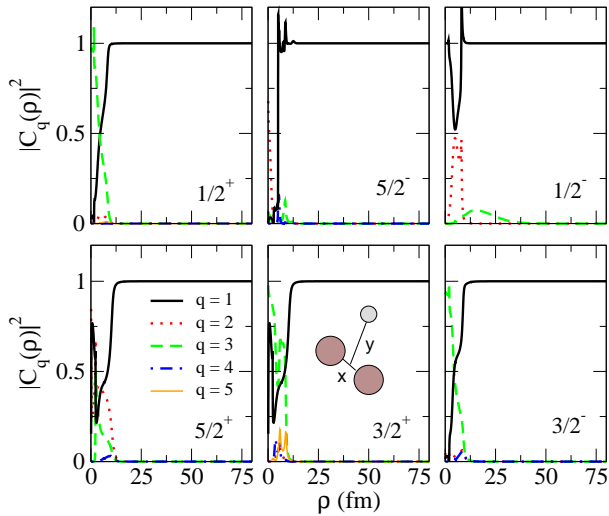}
\end{center}
\vspace*{-0.2cm}
\caption{(Color online) Contribution, as a function of $\rho$, of the
  different partial waves in the first Jacobi set (labeled by q as in
  table~\ref{par}) to the adiabatic eigenfunction
  $\Phi_q(\rho,\Omega)$ related to the ground-state structure of
  $^8$Be at large distance. }
\label{figpar1seq}
\end{figure}

\begin{figure}[th!]
\begin{center}
\vspace*{0.2cm}
\epsfig{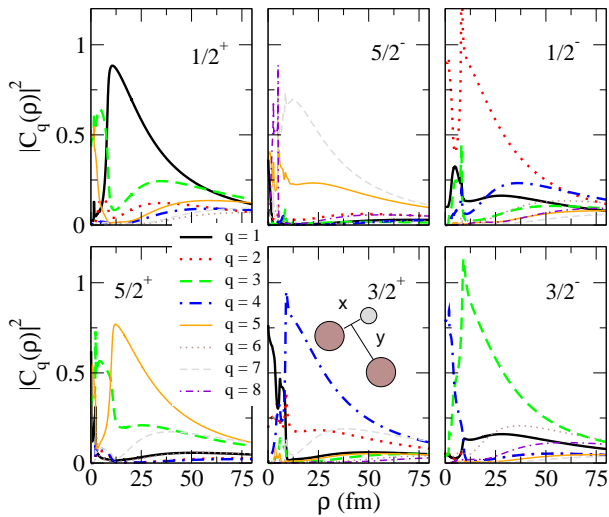}
\end{center}
\vspace*{-0.2cm}
\caption{(Color online) The same as in Fig.~\ref{figpar1seq} but for
  the second Jacobi set. $q$ labels the set of quantum numbers as in
  table~\ref{par2}. }
\label{figpar2seq}
\end{figure}

If this state is populated at large distances, where all couplings to
other adiabatic potentials have vanished, the decay can be described as
sequential via the $^8$Be two-body ground state. Such a decay has a
special role because it is favored by a very low energy with
non-vanishing coupling to other potentials for all the resonances.
Figs.~\ref{figpar1seq} and \ref{figpar2seq} show the partial-wave
decomposition, as a function of the hyperradius, for this adiabatic
component.  

Not surprisingly, the dominant partial wave in the first set of Jacobi
set is the $^8$Be($0^+$) structure for $\rho$ values beyond about
$10$~fm, see Fig.~\ref{figpar1seq}. The same simple structure does not
appear in the second Jacobi coordinates as seen in
Fig.~\ref{figpar2seq}. At short-distances, around 20~fm, one of the
components gives most of the contribution, but this structure is not
maintained at large-distances where we observe a very fragmented
partial-wave decomposition. The reason is of course that the
transformation of the $\ell_x=0$ state in the first Jacobi set into
the second one results in contributions from many different angular
momentum components.

\begin{figure}[th!]
\begin{center}
\vspace*{0.2cm}
\epsfig{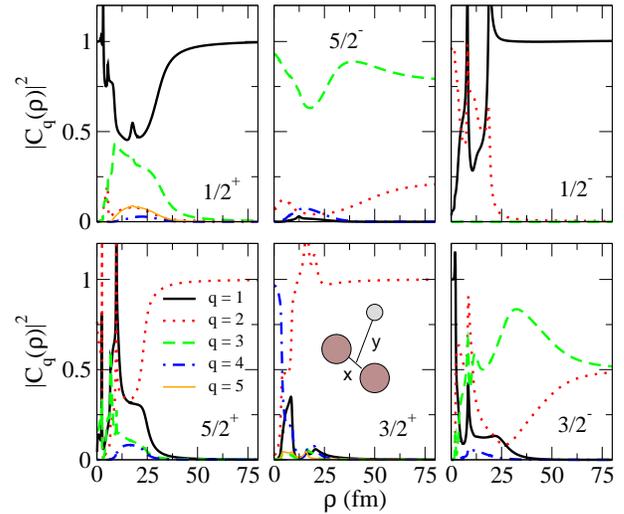}
\end{center}
\vspace*{-0.2cm}
\caption{(Color online) Contribution, as a function of $\rho$, of the different
  partial waves in the first Jacobi set (labeled by q as in
  table~\ref{par}) to the dominant adiabatic eigenfunction
  $\Phi_q(\rho,\Omega)$ different from the one related to the ground
  state structure of $^8$Be at large distance. }
\label{figpar1dir}
\end{figure}

\begin{figure}[th!]
\begin{center}
\vspace*{0.2cm}
\epsfig{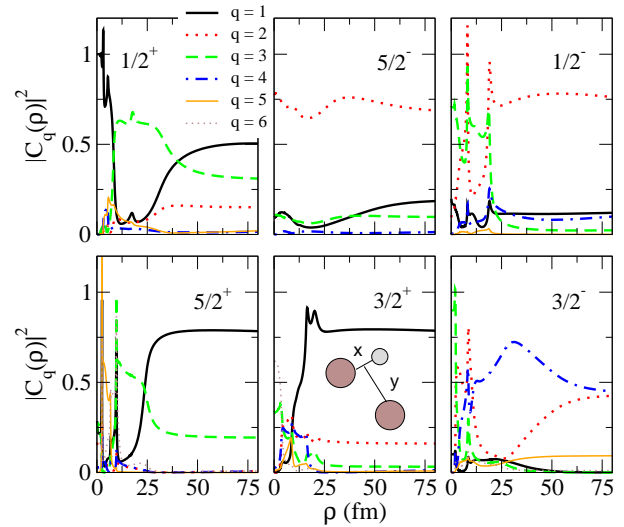}
\end{center}
\vspace*{-0.2cm}
\caption{(Color online) The same as in Fig.~\ref{figpar1dir} but for
  the second Jacobi set.  $q$ labels the set of quantum numbers in
  each component as in table~\ref{par2}.}
\label{figpar2dir}
\end{figure}

The sequential decay via the $^8$Be ground state leads to very simple
momentum distributions derived from the two-body character where
energy and momentum conservation fully determine the final state.  The
remaining part of the decay proceeds through other adiabatic
components. We show in Figs.~\ref{figpar1dir} and \ref{figpar2dir} the
partial-wave decomposition as function of hyperradius for the most
contributing of these other components.

Each of the $J^\pi$ states presents its own features. In all cases the
variation from small to large distances is substantial and sometimes
dramatic.  The structures always converge at large distances.  In the
first Jacobi (Fig.~\ref{figpar1dir}), one of the partial waves absorbs
after convergence almost all the contribution for the resonance of
$1/2^+$, $1/2^-$, $5/2^+$ and $3/2^+$. The corresponding partial waves
have $\ell_x=0,0,2$ and $2$, respectively. They are, therefore, related
to $0^+$ and $2^+$ states in $^8$Be.  In the $5/2^-$ resonance
there are two partial waves contributing significantly, but one of
them is dominating with $\ell_x=2$ and thus related to a $^8$Be($2^+$)
structure \cite{alv08}.  In contrast, the $3/2^-$ resonance has two
components with $\ell_x=2$ that are equally important at large values of
$\rho$.

We show in Fig.~\ref{figpar2dir} the partial-wave decomposition for
these resonances in the second Jacobi system.  The $1/2^+$ state
presents a mixture of three components contributing significantly at
large distances.  This is somewhat analogous to the transform of the
$^8$Be ground state with many partial waves at large distances. Here
we only find three which therefore also emphasizes that even with the
same quantum numbers (all $s$-waves) the structure can be very
different.  The $5/2^-$ and $1/2^-$ states reveal structures where
each converges to quantum numbers identical to those of
$^5$He(p$_{3/2}$) and $^5$He(p$_{1/2}$) respectively.  Again this does
not imply that these are the decay channels, only the quantum numbers
are the same.  In $5/2^+$ and $3/2^+$ the dominant components at large
distance have $\ell_x=0$ as for the $^8$Be ground state but the
behavior differs very much from those of Fig.~\ref{figpar2seq}.  The
last state of $3/2^-$ presents two roughly equal contributions both
with $(\ell_x,\ell_y)=(1,2)$ very much like in the first Jacobi
system. It is like $^8$Be($0^+$) is replaced by
$^5$He($p$-state).

\subsection{Momentum distributions}

The decay mechanisms depend on the resonance properties and they are
conventionally called either sequential via a given two-body structure
or direct decay to the continuum or perhaps a mixture of these
possibilities.  The process is sequential when the measured kinematics
reveals that one particle is first emitted and subsequently the
remaining structure decays into two fragments independent of the
emission of the first particle.  Combinations of such decay channels
form the basis for the $R$-matrix analyses of experimental data
\cite{lan58}. This formulation becomes dubious when the intermediate
two-body structure falls apart on the same time-scale as the first
emission.  The process is then better described as a genuine
three-body decay. This does not prevent analyses in terms of several
two-body decay channels. The two different formulations may still be
completely identical provided the two different sets of basis
functions span the same space. The two formulations merely differ in
the choice of basis \cite{fyn09}.

The present case is in one way rather simple since all the resonances
can decay sequentially via $^8$Be($0^+$), which is a long-lived stable
structure surviving long time after emission of the neutron. Thus the
sequential decay mechanism through this state is not controversial and
easily separated kinematically in experiments.  One of the adiabatic
components is related to the $^8$Be+$n$ structure and approaches the
energy of the $^8$Be($0^+$) resonance.  This component
describes the sequential decay contribution through this channel.
Extension of this picture to sequential decays through $^8$Be($2^+$)
or $^5$He(p$_{3/2}$) is an invitation for difficulties, since these
channels are broad (short-lived) resonance structures, not easily
separated from the background continuum, and furthermore not even
orthogonal contributions. This is more reasonably described as a direct
decay.

\begin{table*}
  \caption{$^9$Be resonance excitation energies, energies above the $\alpha\alpha n$ threshold, and,
for each resonance, estimated amount of computed and observed sequential decay via $^8$Be($0^+$). }
\label{dirseq}
\vspace*{0.2cm}
\begin{tabular}{|ccccccc|}
  \hline
  $J^\pi$  & $E_{\alpha\alpha n}$ (MeV) & $E_{res}$ (MeV) & 
  Theo. (\%) & Exp.(\%) \cite{bro07} & Exp. (\%)\cite{ang99,til04} & 
  Exp. (\%) \cite{bur10} \\  
  \hline
  $\frac{1}{2}^+$ & 1.68 & 0.11 & 100 &           &        & 100 \\
  $\frac{5}{2}^-$ & 2.43 & 0.86 & 3   & $6\pm 1$  & $7\pm 1$&     \\
  $\frac{1}{2}^-$ & 2.82 & 1.25 & 90  & $32\pm 15$& 100    &     \\
  $\frac{5}{2}^+$ & 3.03 & 1.46 & 53  & $46\pm 20$& $87\pm 13$ &\\
  $\frac{3}{2}^+$ & 4.69 & 3.12 & 1   & $16\pm 2$ &     &\\
  $\frac{3}{2}^-$ & 4.22 & 2.65 & 29  &           &     &\\
  \hline
\end{tabular}
\end{table*}

The technique involved is described in \cite{alv08b} where the
large-distance asymptotic behavior of the radial wave functions are
shown to give the ``branching ratio'' for such sequential decay.  We
have calculated this fraction of decay for each of the $^9$Be
resonances. The result is given in table~\ref{dirseq}. The decays of
$\frac{1}{2}^{\pm}$ are found to be predominantly sequential. In
$\frac{5}{2}^+$ both mechanisms are comparable whereas the direct
decays dominate for the other three resonances. The comparison to
measured branching ratios is rather favorable in view of the
uncertainties for broad resonances and the different methods of
extraction.  

The uncertainties are especially emphasized by considering the
$\frac{1}{2}^{-}$ state which often is quoted as predominantly
decaying through the $^8$Be ground state \cite{chr66,che70} in
agreement with our result. This is intuitively appealing since
the alternative channels of $^8$Be($2^+$) and $^5$He($p_{1/2}$) are rather
high-lying.  More recently also contributions through such channels
are extracted from experimental analysis \cite{pre04,bro07} although
given with reservations and uncertainties. Furthermore, the
beta-feeding, the width, and the decay channel are linked together for
broad resonances in data analysis \cite{nym90}.  We conjecture that
the width should be smaller than the measured value in table~\ref{tab}
and the predominant decay channel is through the ground state of
$^8$Be. 

For the sequential channels, the resulting momentum distributions are
easily found, since the first emission immediately provides the energy
of the particle in the three-body center of mass system.  The
following decay is again given by one energy in the center of mass
system of the remaining two particles.

The momentum distributions for direct decays into the continuum can
now be found by excluding the sequential contribution, that is the
part of the wave function residing in the $^8$Be ground state at large
distance. Again we have to calculate, as accurately as possible, the
large-distance asymptotics of the wave function.  The technique,
described in \cite{fed04,gar06}, is based on finding the Zeldovic
regularized Fourier transform of the coordinate state wave resonance
function.  The result is directly comparable to measured
distributions. It is worth emphasizing again that the only link from the
asymptotic, measurable distribution, to the small-distance structure
is via theoretical models \cite{alv08}.

\begin{figure*}
\begin{center}
\vspace*{-.2cm}
\epsfig{file=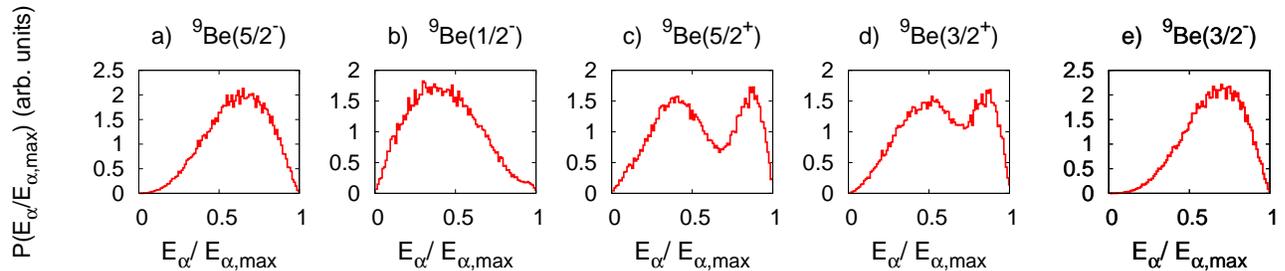,scale=0.248,angle=-90}
\end{center}
\vspace*{-0.5cm}
\caption{(Color online) $\alpha$-particle energy distributions for a)
  the $5/2^-$-resonance of $^9$Be at 2.43~MeV of excitation energy (or
  0.86~MeV above the $\alpha\alpha n$ threshold), b) the
  $1/2^-$-resonance of $^9$Be at 2.82~MeV of excitation energy (or
  1.25~MeV above the $\alpha\alpha n$ threshold), c) the
  $5/2^+$-resonance of $^9$Be at 3.03~MeV of excitation energy (or
  1.46~MeV above the $\alpha\alpha n$ threshold), d) the
  $3/2^+$-resonance of $^9$Be at 4.69~MeV of excitation energy (or
  3.12~MeV above the $\alpha\alpha n$ threshold), e) the
  $3/2^-$-resonance of $^9$Be at 4.22~MeV of excitation energy (or
  2.65~MeV above the $\alpha\alpha n$ threshold). The energies are
  divided by the maximum possible, i.e. $5/9E_{res}$. The sequential
  decay via $^8$Be($0^+$) has been removed in all the cases. }
\label{figendis_a}
\end{figure*}

\begin{figure*}
\begin{center}
\vspace*{-.2cm}
\epsfig{file=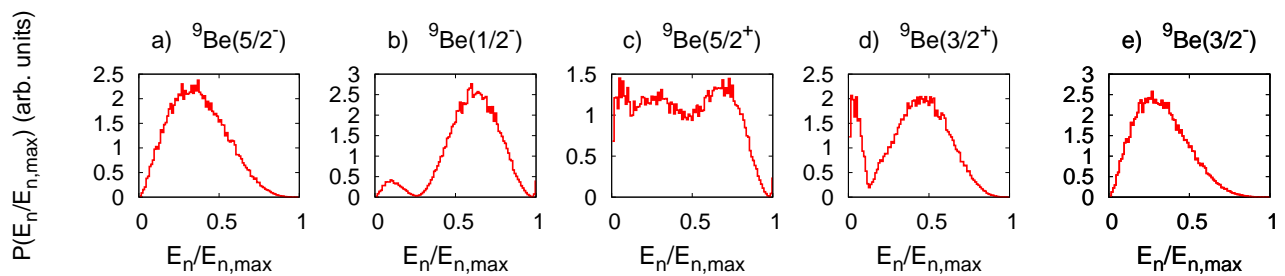,scale=0.248,angle=-90}
\end{center}
\vspace*{-0.5cm}
\caption{(Color online) The same as Fig.~\ref{figendis_a} for the
  neutron energy distributions.  The energies are divided by the
  maximum possible, i.e. $8/9E_{res}$.}
\label{figendis_n}
\end{figure*}

We compute the distributions by Monte Carlo simulation. We first
generate randomly a large number of events, each of them consisting of
three four-momenta relative to our three decaying fragments. The sum of
their center-of-mass energies must equal the resonance energy. The
weight of each set of momenta is the absolute-squared wave-function at
large distance.  The resulting energy distributions are shown in
Figs.~\ref{figendis_a} and \ref{figendis_n} for $\alpha$-particles and
neutrons, respectively.  We give the energies in units of their
maximum values for each case, i.e. $5/9E_{res}$ for the $\alpha$'s and
$8/9E_{res}$ for the neutrons.

\begin{figure*}
\begin{center}
\vspace*{-.2cm}
\epsfig{file=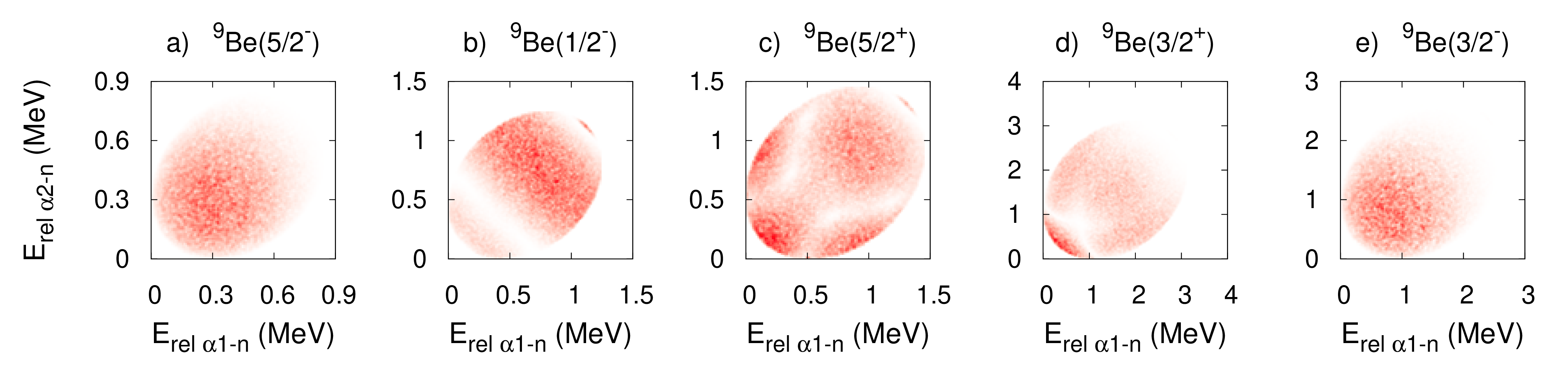,scale=0.248,angle=-90}
\end{center}
\vspace*{-0.5cm}
\caption{(Color online) Dalitz plots for a) the $5/2^-$-resonance of
  $^9$Be at 2.43~MeV of excitation energy (or 0.86~MeV above the
  $\alpha\alpha n$ threshold), b) the $1/2^-$-resonance of $^9$Be at
  2.82~MeV of excitation energy (or 1.25~MeV above the $\alpha\alpha
  n$ threshold), c) the $5/2^+$-resonance of $^9$Be at 3.03~MeV of
  excitation energy (or 1.46~MeV above the $\alpha\alpha n$
  threshold), d) the $3/2^+$-resonance of $^9$Be at 4.69~MeV of
  excitation energy (or 3.12~MeV above the $\alpha\alpha n$
  threshold), e) the $3/2^-$-resonance of $^9$Be at 5.59~MeV of
  excitation energy (or 4.02~MeV above the $\alpha\alpha n$
  threshold). We plot on the axis the $\alpha$-n relative energy in
  MeV. The sequential decay via $^8$Be($0^+$) has been removed in all
  the cases. }
\label{figdal}
\end{figure*}

The distributions all necessarily have peaks, since they start with
zero and return again to zero at maximum energy. However, they can
have more than one peak, and each of them has an individual position
and width. The $5/2^-$ and $3/2^-$ resonances are smooth with one peak
for both neutrons and $\alpha$-particles. In both cases, the neutron
energies peak below and the $\alpha$-particle above half of their
respective maximum values.  This means a tendency to emit
$\alpha$-particles in essentially opposite directions while leaving
the remaining neutron in the middle with relatively little
energy. This is only a tendency and the full distributions require
detailed computations. Still it is indicative for this part of the
process.

For the $1/2^-$ decay the $\alpha$-particle show up with a broad
distribution on the low-energy side whereas the neutron appears on the
high-energy side. This resembles the sequential decay through the
$^8$Be ground state where $\alpha$-particles end up with only little
energy. However, the present decay has to proceed through an
orthogonal adiabatic potential which reveals itself by the low-energy
node in the distribution of the neutron energy.

The $3/2^+$ and $5/2^+$ resonances both produce neutrons and
$\alpha$'s with tendencies to be respectively on high and low-energy
sides like for the $5/2^-$ and $3/2^-$ resonances. However, in the
positive parity cases additional peaks appear in both distributions,
again a signal of an excited state.  These cases are otherwise not
very similar and the distributions are very broad each extending
across from high to low-energy side and vice versa.

These distributions can be suggestive and deceiving. The momenta are
distributed among all the three particles which is the reason for the
continuous distributions in the first place. However, this also means
that a kinematically complete description for a given conserved total
resonance energy requires energies of two particles at the same time.
This information is contained in the two-dimensional energy
correlations known as Dalitz plots which were introduced by
R.H. Dalitz in 1953 to study decays of K-mesons \cite{dal53}.  These
correlation diagrams provide an excellent tool for studying the
dynamics of three-body decays.  The technique has recently been picked
up and applied in studies of nuclear fragmentation processes
\cite{fyn03}. In simple two-body decays the angular
distribution of the emitted particles carries the signature of
decaying angular momentum and parity. The Dalitz plots are
generalizations to three-body decays and it is natural to use the
plots in attempts of experimentally assigning spin and parity to the
decaying resonances, \cite{fyn09}.

To establish the connection to measured distributions we computed
Dalitz plots for $\alpha$-particles and neutrons after the decay of
$^9$Be resonances. We use the same Monte Carlo technique as for the
individual particles energy distributions. To facilitate comparison
with the experimental results from \cite{bro07} we plot the $\alpha-n$
relative energies on the $x$ and $y$-axes, i.e.
\begin{equation}
E_{\alpha-n}= \frac{|\bm{p}_{\alpha-n}|^2}{2\mu_{\alpha n}} \;,\,\,
\bm{p}_{\alpha-n} = \frac{\mu_{\alpha n}}{m_{\alpha}}\bm{p}_\alpha -
\frac{\mu_{\alpha n}}{m_{n}}\bm{p}_n \;,
\end{equation}
where $\bm{p}_{\alpha-n}$ is the relative momentum.  The results are
shown in Fig.~\ref{figdal}.  We first observe that all the
distributions are symmetric with respect to interchange of the axes.
This is necessary and reflects that the wave functions are symmetric
for the identical bosonic $\alpha$-particles.

The graphs corresponding to $\frac{5}{2}^-$ and $\frac{3}{2}^-$, are
very similar to each other.  None of them exhibits any points or
regions of zero probability, except on the confining envelope defined
by energy conservation.  This means that symmetry, angular momentum
and parity of these structures, $(\ell_x,\ell_y)=(1,2),(2,1)$ (tables
\ref{par} and \ref{par2}), allow emission in all directions and with
all energy partitions.

The probability increases towards higher $\alpha$ energies, which
corresponds to smaller relative energies since the neutron is much
lighter than the $\alpha$-particle. This is due to the Coulomb
repulsion and the tendency to choose a decay path where the neutron is
left in the middle as observed in the one-dimensional energy
distributions, see Figs.~\ref{figendis_a} and \ref{figendis_n}.
Distributions for both states compare well with the experimental plots
from \cite{bro07}.  Moreover, the $\frac{5}{2}^-$ distribution is very
similar to the measured ones published in \cite{pap07}, and
investigated theoretically in detail in \cite{alv08}.

The distributions for $\frac{1}{2}^-$, $\frac{5}{2}^+$,
$\frac{3}{2}^+$ exhibit much more structure and all have zero
probability regions as reflected in the nodes or minima of the
one-dimensional distributions. For $\frac{1}{2}^-$ we find a striking
similarity with the measured distribution in \cite{bro07} at the
excitation energy window at $2.8$~MeV. The very low probability bands
at lower and higher relative energies are found both places.  The
different projection on the neutron energy axis in
Figs.~\ref{figendis_n} resulted in a node at small neutron energy,
presumably corresponding to a cut along low energy small probability
region in Fig.~\ref{figdal}.

For both $\frac{5}{2}^+$ and $\frac{3}{2}^+$ the computed
distributions have lots of structure whereas the measurements show
more smooth distributions without much resemblance to calculations.
The explanation for these discrepancies is still to be found.

\section{Summary and conclusions}

The hyperspherical adiabatic expansion method, combined with complex
scaling, is used to compute the energies and widths of $^9$Be
low-lying resonances. We describe them as three-cluster resonances
($\alpha \alpha n$). Realistic short-range nuclear interactions as
well as Coulomb interactions are included in the computations.  To
reach high accuracy we use a large hyperharmonic basis for each
angular eigenfunction, accurate large-distances, outgoing waves of
radial wave functions and, if possible, the correct energy of the
three-body resonance obtained by tuning the three-body potential.

We find one bound state ($\frac{3}{2}^-$) and six resonances 
below 6~MeV of excitation energy in agreement with experimental
information. Spins and parities of the resonances are
$\frac{1}{2}^\pm$, $\frac{3}{2}^\pm$ and $\frac{5}{2}^\pm$.  The
small-distance properties of the adiabatic potentials determine
energies, while barriers at intermediate distances are crucial for the
widths, and the large-distance structure of the resonances are
decisive for the momentum partition between the three particles in the
final state after decay.

The structure of the resonances are obtained as different combinations
of angular momenta of the two-body subsystems. The configurations are
determined by the interactions leading to observed low-lying
resonances of the subsystems, i.e. $0^+,2^+$ for $^8$Be and $p_{3/2}$
$p_{1/2}$ for $^5$He. The detailed configurations of the three-body
resonances are extracted, their energies fine-tuned via the three-body
potential, and their widths computed. 

We compute the possibly substantial dynamic evolution of the
resonances as functions of hyperradius.  The large-distance asymptotic
structures are via Fourier transformation directly related to the
momentum distributions of the fragments after the three-body decay.
We determine the fraction decaying via the ground state of $^8$Be in a
sequential decay. The agreement with measurements is rather good in
view of the uncertainties related to broad resonances and different
theoretical and experimental definitions and methods.

The remaining part is described as direct decay to the three-body
continuum. We present the computed momentum distributions of neutrons
and $\alpha$-particles for each of the resonances.  These observable
distributions are results of the dynamic evolution, and open to
experimental tests.  We compare with the available data, and find
remarkable similarities except for the $5/2^+,3/2^+$ resonances where
the theory gives much more structure than found in the energy windows
selected in the experiments.

\section*{Acknowledgments}
This work was partly supported by funds provided by DGI of MEC (Spain)
under contract No.  FIS2008-01301 and the Spanish Consolider-Ingenio
programme CPAN (CSD2007-00042).  R.A.R. acknowledges support by
Ministerio de Ciencia e Innovaci\'on (Spain) under the ``Juan de la
Cierva'' programme. We have benefited from continuous discussions with
H. Fynbo and K.Riisager.

\end{document}